\documentclass[aps,pra,twocolumn,notitlepage]{revtex4-1}

\usepackage{graphicx}

\begin{document}

\title{An exploration of the limits of control using quantum superpositions}

\author{Holger F. Hofmann}
\affiliation{
Graduate School of Advanced Sciences of Matter, Hiroshima University,
Kagamiyama 1-3-1, Higashi Hiroshima 739-8530, Japan}

\begin{abstract}
Quantum interferences between non-orthogonal states are the best approximation of a joint realization of
the non-commuting physical properties represented by the two states. As I have shown recently, such
interferences can be used to demonstrate that quantum physics deviates from classical causality in the limit
of small action. Here, I point out that this proof constitutes a failure of the principle of least action and
consider possible implications for our understanding of fundamental physics. 
\end{abstract}


\maketitle

\section{Introduction}

Although quantum information research has given us a wide variety of methods to prepare and manipulate the quantum states of physical systems, there is still no satisfactory explanation of the mechanisms that govern quantum processes. Several recent attempts to address the problem have focused on the role of causality in quantum computation and in similar quantum networks \cite{Cha15,Cos16,Cav18}. However, this logical approach to causality seems to overlook the fundamental role that causality and determinism already plays in the original formulation of quantum mechanics, where the continuous time evolution of a state is understood as the quantum mechanical limit of classical equations of motion. It may therefore be necessary to investigate the quantum mechanical modifications of classical relations between observable properties in more detail. In this context, it may also be important to remember that the uncertainty principle usually prevents us from establishing a relation between more than two non-commuting observables, a fact that makes it very difficult to argue about causality in quantum systems. This is why the application of weak measurements to quantum paradoxes and to measurement uncertainties has provided us with such a large number of new and potentially confusing results \cite{Res04,Yok09,Gog11,Suz12,Den14,Hof15,Hal04,Has12,Roz12,Bae13}. To understand all of these results in a wider context, we should focus on the fact that weak measurements and weak values establish a particularly close relation between quantum coherence and causality \cite{Lun12,Hof12a,Hof14,Nii17}. The mathematical formalism itself suggest that the quantum coherence between two non-orthogonal states corresponds to a logical ``AND,'' as can be seen in the analysis of the quantum statistics of optimal cloning \cite{Hof12b,The17}. Weak measurements isolate the quantum coherence between the weakly measured observable and the post-selected observable, resulting in an effective joint measurement of two non-commuting properties. In a time-reversal of this process, it is possible to achieve a maximal statistical contribution of two non-orthogonal states by preparing a constructively interfering superposition of the two. It is then possible to show that the interference pattern represents a modification of causality, as I have done for the case of particle propagation in \cite{Hof17}.  
 
In this presentation, I will explain the principles that result in the violation of classical causality demonstrated in \cite{Hof17}. By relating the results to our recent work on the role of the action in quantum interference \cite{Hof16,Hib18} it can be shown that quantum mechanics replaces the principle of least action with non-classical correlations represented by an interference pattern, where the action determines the phase of the interference fringes in such a way that the majority of the probability is accumulated near (but not at) the point of least action. Causality and the associated possibilities of control thus originate from the dynamical relation between physical properties expressed by the action and not from the direct access to physical properties mistakenly envisioned by ``realist'' models.

\section{Determinism and control}
As discussed in \cite{Hof17}, the basic logic of deterministic control is best illustrated by Newton's first law. If initial position and momentum were known with precision, we could identify all intermediate positions, since they would have to lie on a straight line given by these initial conditions. Unfortunately, it is difficult to explain how quantum mechanics modifies this logic, since we are fundamentally unable to express a joint control of position and momentum within the established formalism of quantum theory. The original justification for this failure to address the problem of causality is that there is no practical way of exceeding the uncertainty limits, so we should not expect a theoretical explanation for something that cannot be observed. However, the quantum formalism introduces a completely new way of describing causality in the form of quantum coherence, and we should expect a more detailed explanation of how and why the classical notion of causality emerges within this description.

\cite{Hof17} introduces a new method to address this problem by using quantum interferences between initial conditions $A$ and $B$ to identify a minimal contribution of ``$A$ AND $B$''. For pure state representations with real-valued inner products $\langle B \mid A \rangle=\langle A \mid B \rangle$, an equal superposition is given by
\begin{equation}
\label{eq:state}
\mid \psi \rangle = \frac{1}{\sqrt{2(1+\langle B \mid A \rangle)}}\left(\mid A \rangle + \mid B \rangle \right).
\end{equation}
In this superposition, constructive interference enhances the probabilities of finding $A$ or $B$ in a corresponding measurement. Specifically, the sum of the measurement probabilities $P(A)$ and $P(B)$ exceeds one by a probability equal to the quantum overlap,
\begin{equation}
\label{eq:limit}
P(A)+P(B)-1 = \langle B \mid A \rangle.
\end{equation}
Even though $A$ and $B$ cannot be measured jointly, constructive interference suggests that any possible joint distribution of $A$ and $B$ would have to include a minimal joint probability of $\langle B \mid A \rangle$ to satisfy Eq.(\ref{eq:limit}). In this sense, quantum interference allows us to realize a non-trivial level of joint control over the incompatible conditions $A$ and $B$.  

\section{Contributions of the interference term}
We can now consider the effects of the simultaneous control of $\mid A \rangle$ and $\mid B \rangle$ on an observable $\hat{x}$. In quantum mechanics, the relation between $\mid A \rangle$ and $\mid x \rangle$ and the relation between $\mid B \rangle$ and $\mid x \rangle$ are given by the inner products of the state vectors. It is possible to identify situations where $\hat{x}$ depends on both $\mid A \rangle$ and $\mid B \rangle$ in a symmetric fashion, such that 
\begin{equation}
|\langle x \mid A \rangle| = |\langle x \mid B \rangle|.
\end{equation}
The difference between the two states is then expressed completely by the quantum phases of the inner products. Specifically, the interference pattern produced by the state in Eq.(\ref{eq:state}) can be expressed in terms of the phase difference $S(x)/\hbar$ between $\langle x \mid A \rangle$ and $\langle x \mid B \rangle$,
\begin{equation}
|\langle x \mid \psi \rangle|^2 = \frac{1}{(1+\langle B \mid A \rangle)}\left(1+\cos(S(x)/\hbar)\right)|\langle x \mid A \rangle|^2. 
\end{equation}
Normalization requires that the oscillations associated with quantum interferences between $A$ and $B$ do not average out completely. Since the probability distribution $|\langle x \mid A \rangle|^2$ is normalized, we find that
\begin{equation}
\int \cos(S(x)/\hbar)|\langle x \mid A \rangle|^2 \;dx = \langle B \mid A \rangle.
\end{equation}
This relation must be satisfied even if the probability distribution $|\langle x \mid A \rangle|^2$ is much wider than the interference fringes and changes only very little during one period of oscillation associated with an action difference of $2\pi \hbar$. The reason for a non-zero contribution of the integral can then be identified with points of stationary action, where the $x$-derivative of the quantum phase drops to zero and changes sign.
\begin{equation}
\label{eq:leastact}
\frac{\partial}{\partial x} S(x) \big|_{x=\mu} = 0.
\end{equation}
If there is only one such point at $x=\mu$, the integral can be approximated by a Fresnel integral of the form
\begin{equation}
\int \cos(\gamma (x-\mu)^2-\pi/4)\, dx = \sqrt{\frac{\pi}{\gamma}} 
\end{equation}
and the corresponding relation between the inner products can be given as
\begin{equation}
\label{eq:relate}
\sqrt{\frac{\pi}{\gamma}} |\langle x=\mu \mid A \rangle|^2 \approx \langle B \mid A \rangle,
\end{equation}
where
\begin{equation}
\gamma = \frac{1}{2 \hbar} \frac{\partial^2}{\partial x^2} S(x) \big|_{x=\mu}.
\end{equation}
As we shall see in the following, this relation between the overlap $\langle B \mid A \rangle$, the probability density at $x=\mu$, and the second derivative of the action in $x$ at $x=\mu$ represent a quantum mechanical correction in the laws of causality that relate $A$ and $B$ to $x$.  

\section{Action and uncertainty}

The key insight from my previous analysis of quantum coherence is that the complex phases of quantum mechanics all represent deterministic relations between physical properties, where the complex phase itself is the action that describes the transformation dynamics associated with these properties \cite{Hof14,Hof16,Hib18}. It might be worth noting that this action is closely related to the quantum action of Schwinger's variational principle, as shown by Dressel and coworkers in \cite{Dre14}. Unfortunately, the traditional approaches to principles of least action in quantum mechanics tend to ignore or downplay the problem that quantum interferences do not reproduce classical determinism. Here, I would like to give a compact explanation of this problem using the interference between $\mid A \rangle$ and $\mid B \rangle$ in $x$. 

As I already explained in the previous section, the non-zero contribution to the constructive interference $\langle B \mid A \rangle$ that represent the quantum mechanical equivalent of ``$A$ AND $B$'' originates from $x$-values close to $x=\mu$, where the action $S(x)$ has its minimum. In the classical limit, this condition is misinterpreted as a determinism of quantities, where $\mu=x(A,B)$ is understood as a mathematical function of two initial conditions, $A$ and $B$. Importantly, this misinterpretation is completely consistent with our actual experience of physics, making it very tempting to believe that it is somehow proven by the sum of our experience. However, there is no evidence that ``numbers are physical,'' making it a fallacy to base interpretational arguments on the assumption that it is somehow necessary to think so. The details of quantum mechanics suggest that causality is not given by a continuity of numerical quantities, but instead originates from the quantum interference effects associated with the action. The difference appears in the details, particularly in the replacement of precise values with interference patterns. Experimental evidence for the difference between the (classical) principle of least action and quantum causality can be obtained by using the spread of the interference pattern, which shows that the probability of ``$A$ AND $B$'' exceeds the probability of the least action value of $x=\mu$. This is the essence of the proposal in \cite{Hof17}, where the main complication is the accommodation of uncertainties in the two states representing the initial conditions. 

In quantum mechanics, the failure of the principle of least action widens the range of $x$-values contributing to the overlap $\langle B \mid A \rangle$. If the principle of least action was valid, each $x$-value would correspond to a specific combination of $A$ and $B$. Hence, the probability distribution $|\langle x \mid A \rangle|^2$ could be converted into a distribution of values for the second initial condition, where the specific value of $B$ was found at $x=\mu$. Clearly, the finite statistical overlap of $A$ and $B$ given by $|\langle B \mid A \rangle|^2$ is only possible if $B$ can be represented by a finite range of $x$-values, requiring a statistical uncertainty of 
\begin{equation}
\label{eq:class}
\delta x(A \;\mbox{AND}\; B) = \frac{|\langle B \mid A \rangle|^2}{|\langle x=\mu \mid A \rangle|^2}.
\end{equation}
This classical uncertainty limit is merely a consequence of the practical problem of state preparation. If both $A$ and $B$ are determined by continuous variables, their eigenstates will automatically have infinite uncertainties and a mutual overlap of zero, making it necessary to allow for a non-vanishing uncertainty in the definition of the quantitative conditions. The states $\mid A \rangle$ and $\mid B \rangle$ will therefore not be eigenstates of the initial conditions, and their overlap provides a measure of the uncertainties that enter into the relation with $x$. 

We can now directly compare the classical broadening of the deterministic relation between $A$, $B$ and $x$ with the quantum mechanical broadening of the quantum interference pattern described by Eq.(\ref{eq:relate}). In the quantum mechanical limit, contributions to the overlap $\langle B \mid A \rangle$ have a complex phase associated with the action $S(x)$, which is $-\pi/4$ at $x=\mu$ and rises quadratically from this minimum. The value of the cosine function contributions to the overlap actually increases as $x$ moves away from $\mu$, reaching a maximum when the phase passes through zero and dropping back to the original value at a phase of $\pi/4$. The quantum broadening of the interference pattern can therefore be defined as the difference between $x=\mu$ and the $x$-value at which the quantum phase is $\pi/4$,
\begin{equation}
\label{eq:quant}
\delta x(\mbox{quantum}) = \sqrt{\frac{\pi}{2 \gamma}} = \frac{\langle B \mid A \rangle}{|\langle x=\mu \mid A \rangle|^2}.
\end{equation}
Comparison with the statistical uncertainty in Eq.(\ref{eq:class}) which represents the experimental problem of generating initial conditions with a finite overlap shows that the quantum broadening of causality is always larger than the uncertainty of control by a factor that is equal to the overlap between the quantum states,
\begin{equation}
\label{eq:compare}
\delta x(\mbox{quantum}) = \frac{1}{\langle B \mid A \rangle} \delta x(A \;\mbox{AND}\; B).
\end{equation}
Quantum interference thus describes a fundamental modification of the physics of causality \cite{Hof14,Hof17}. Importantly, this modification reveals that all classical formulations of causality are merely approximations of quantum coherent effects. Ultimately, there is no better simultaneous control of $A$ and $B$ than constructive quantum interference between states with low uncertainties in one of the two properties that we wish to control. In that scenario, classical causality would suggest that  
\begin{equation}
|\langle x=\mu \mid \psi \rangle|^2  \delta x(A \;\mbox{AND}\; B) \geq P(A \;\mbox{AND}\; B).
\end{equation}
As shown in \cite{Hof17}, it is possible to violate this inequality, demonstrating the failure of classical causality relations associated with the principle of least action. In the present context, Eq.(\ref{eq:limit}) shows that 
$P(A \;\mbox{AND}\; B)\geq \langle B \mid A \rangle$, so that classical causality requires
\begin{equation}
|\langle x=\mu \mid \psi \rangle|^2  \delta x(\mbox{quantum}) \geq 1.
\end{equation}
It is actually easy to see that this condition can only be satisfied if the distribution of $x$-values in $\mid \psi \rangle$ is narrower than the quantum broadening of causality given by $\delta x(\mbox{quantum})$. However, this condition is only met if both $A$ and $B$ independently determine $x$ with a precision greater than $\delta x(\mbox{quantum})$. Whenever $x$ appears to depend equally on $A$ and on $B$, quantum physics requires causality relations that are based on quantum interference effects, and these causality relations necessarily supersede the classical description of causality as a continuous internal reality of the system.


\section{Conclusions}

It is impossible to provide a scientific definition of reality without referring to the available mechanisms of control. In quantum physics, these mechanisms of control are limited by fundamental laws that modify the causality relations between different physical properties of a system. In my previous research, I have shown that all fundamental causality relations can be expressed by quantum phases, and that these causality relations describe and explain the contextuality of quantum measurements \cite{Suz12,Hof14,Hof15,Hof16,Hib18}. Importantly, the replacement of least action by quantum interference changes the meaning of physical reality in a non-trivial manner. If $x$ is a consequence of $A$ and $B$, the maximal level of control is represented by constructive interferences between the two conditions, and the control of $x$ appears in the interference pattern of the two conditions. This interference pattern is necessarily spread out over a wide range of $x$-values, with positive and negative contributions cancelling each other except in a central region around the least action value of $x=\mu$. Even if the high frequency oscillations are ignored, the region around the least action point at $x=\mu$ where quantum phases are still smaller than $\pi/4$ is much wider than classical causality would allow, revealing a quantitative difference between the classical control of $x$ by $A$ and $B$ and the quantum mechanics of control by interference. 

The demonstration that quantum interference replaces the classical notion of causality expressed by quantitative relations of the form $\mu=x(A,B)$ indicates that the assumption of a continuous reality is redundant because its role as a mediator of causality is taken over by quantum phases. Our sense of objective reality does not require a mathematical descriptions of ``things in themselves.'' Instead, it is entirely sufficient to have a consistent description of all the possible appearances of objects in their physical interactions. What needs to be understood is that the quantum phases of Hilbert space represent the actual process of change in time that is associated with these physical interactions \cite{Hib18}. It is a fallacy to believe that the continuity of causality implies a corresponding continuity of reality, since realities only emerge in the form of forces described by interactions. In these interactions, the internal causality of the system is disturbed by external conditions that represents the physics of control. Ultimately, we need to address the problem that we still seem to think of this control mechanism as ``classical,'' even though the concepts of state preparation and measurement used in quantum physics are quite different from the rather vague notions that would be intuitively associated with them in any classical context. The present summary is intended to provide a new perspective on this problem by pointing a way towards a more detailed understanding of the essential relation between causality and quantum coherence in the microscopic limit of control.  

\vspace{0.2cm}
\noindent
This work was supported by JSPS KAKENHI Grant Number 26220712 and by CREST, Japan Science and Technology Agency.


\end{document}